\documentclass[3p,twocolumn]{elsarticle}
\usepackage{amsmath}
\usepackage{amsthm}
\usepackage{amssymb}
\usepackage{geometry}                 
\usepackage{graphicx}
\usepackage[utf8]{inputenc}
\usepackage[english]{babel}

\begin{document}
\title{Closing the hierarchy for non-Markovian magnetization dynamics}

\author[Dam,Univ]{J. Tranchida}
\ead{julien.tranchida@cea.fr}
\author[Dam]{P. Thibaudeau}
\ead{pascal.thibaudeau@cea.fr}
\author[Univ]{S. Nicolis}
\ead{stam.nicolis@lmpt.univ-tours.fr}

\address[Dam]{CEA/DAM/Le Ripault, BP 16, F-37260, Monts, France}
\address[Univ]{CNRS-Laboratoire de Mathématiques et Physique Théorique (UMR 7350), Fédération de Recherche "Denis Poisson" (FR2964), Département de Physique, Université de Tours, Parc de Grandmont, F-37200, Tours, France}

\date{\today}   

\begin{abstract}

We propose a stochastic approach for the description of the time evolution of the magnetization of nanomagnets, that interpolates between the Landau--Lifshitz--Gilbert and the Landau--Lifshitz--Bloch approximations, by varying the strength of the noise. In addition, we take into account the autocorrelation time of the noise and explore the consequences, when it is finite, on the scale of the response of the magnetization, i.e. when it may be described as colored, rather than white, noise and non-Markovian features become relevant. We close the hierarchy for the moments of the magnetization, by introducing a suitable truncation scheme, whose validity is tested by direct numerical solution of the moment equations and compared to the average deduced from a numerical solution of the corresponding stochastic Langevin equation. In this way we establish a general framework, that allows both coarse-graining simulations and faster calculations beyond the truncation approximation used here. 

\end{abstract}

\maketitle
\setcounter{page}{1}
\setcounter{section}{0}

\section{Introduction}
At atomistic length-scales, relaxation processes towards equilibrium of the magnetization in magnetic systems  rely on complex interactions between spin, electron and lattice subsystems \cite{Suhl:2007qa,gurevich1996magnetization,Radu2011}. 
The time-scales for these processes are extremely short ($\tau\approx100\pm80$ps for a ferromagnetic gadolinium system \cite{vaterlaus1991spin}), 
so a  first approach for their description  was to perform stochastic simulations of interacting spins while neglecting  the memory effects for the noise (Markov's hypothesis) \cite{Brown-Jr:1963tp,coffey2012thermal}.
However, recent experimental breakthroughs are pushing the time-scales that can be probed for the magnetization dynamics to pico- and even femtosecond resolution\cite{kimel2005ultrafast}. 
Furthermore, a self--consistent approach for  the short-range exchange interaction, that is responsible for local spin alignment, indicates that it would be equivalent to a very high value of the local magnetic field 
(more than $80~\rm{MAm}^{-1}$). To take into account this contribution in simulations, it seems to  impose time-steps at least one order of magnitude shorter than all the physical time scales (dynamics, relaxation, etc.) of the system. 
Under these conditions, numerical simulations of the stochastic effects in such magnetic systems are raising  the  question of the validity of the  Markovian assumption that is commonly used, that is to have short memory in the sense that the correlation time is very short~\cite{atxitia2009ultrafast}.
In order to test this assumption, by taking memory effects explicitly into account, a colored-form for the noise has to be considered. 
In spin systems, the noise is represented by $\vec{\tilde{\omega}}$ as a random vector whose the components, $\tilde{\omega}_i$, are  Gaussian random variables with zero mean and a the finite correlation time, which is encoded in its two-point function as follows:
\begin{equation}
\langle \tilde{\omega}_i(t) \tilde{\omega}_j(t') \rangle=\frac{D}{\tau}\delta_{ij} \exp \left(-\frac{\lvert t-t' \rvert}{\tau} \right)
\label{Colorednoise}
\end{equation}
where $D$ is the amplitude of the noise, and $\tau$ is the correlation time.
We remark that $\lim\limits_{\tau \to 0} \frac{1}{\tau}\exp \left(-\frac{\lvert t-t' \rvert}{\tau} \right)=\delta(t-t')$ which allows the white-noise expression for the correlator to be recovered in this limit.

To compute the moments of the magnetization, subject to such noise, we adapt 
 the formalism developed by Shapiro and Loginov \cite{shapiro1978formulae}  to the magnetization dynamics.
A new hierarchy of equations for the moments is then deduced and appropriate 
closure relations allow us to solve them directly. The solution is compared to numerical simulations of the corresponding stochastic Langevin equations, that are performed in the vicinity of the white noise limit. 

\section{Moment equations and closure of the hierarchy}
The Einstein summation convention is adopted, with  Latin indices standing for vector components, $\epsilon_{ijk}$ is the anti-symmetric Levi-Civita pseudo-tensor and $s_i$ are the components of the normalized magnetization vector ($\left|\vec{s} \right|=1$). The stochastic Landau-Lifshitz-Gilbert (sLLG) equation can be written as follows:
\begin{eqnarray}
\frac{\partial s_i}{\partial t}&=&\frac{1}{1+\lambda^2} \epsilon_{ijk} s_k \left[\omega_j + \tilde{\omega}_j \right.\nonumber \\
                              ~&~&-\left.\lambda \left(\epsilon_{jlm} (\omega_l+ \tilde{\omega}_l) s_m \right) \right]
\label{LLG0}
\end{eqnarray}
Here  $\vec{\omega}$ is the  effective precession frequency of the magnetization, that is assumed to be constant, $\lambda$ is the damping coefficient, and $\vec{\tilde{\omega}}$ represents the stochastic noise contribution, whose components are drawn from a Gaussian distribution with zero mean and a colored form for the second order correlator~eq.~(\ref{Colorednoise}). 
This is the fundamental equation for the dynamics of a small magnet in contact with a stochastic bath, which we shall take to be thermal henceforth. It has the form of a Langevin equation, with {\em multiplicative} noise~\cite{justin1989quantum}. 

It can be simplified by keeping only one random torque on the right-hand side:
\begin{equation}
\frac{\partial s_i}{\partial t}=\frac{1}{1+\lambda^2} \epsilon_{ijk} s_k \left[\omega_j + \tilde{\omega}_j -\lambda\epsilon_{jlm} \omega_l s_m \right]
\label{LLG}
\end{equation}
Indeed, up to a renormalization of the noise, it can be shown that  this equation generates  a stochastic dynamics that is equivalent to that of eq.~(\ref{LLG0})\cite{mayergoyz2009nonlinear}.

These microscopic degrees of freedom give rise to average quantities, that probe statistical ensemble behavior. These can be operationally defined by taking statistical averages over the noise \cite{justin1989quantum} of eq.~(\ref{LLG}). 
Denoting by  $\langle . \rangle$ this average, we have:
\begin{eqnarray}
\frac{\partial \langle s_i\rangle}{\partial t}
&=&\frac{1}{1+\lambda^2} \left[ \epsilon_{ijk}\omega_j \langle s_k\rangle+ \epsilon_{ijk} \langle\tilde{\omega}_j s_k\rangle\right.\nonumber\\
&~&\left.-\lambda \epsilon_{ijk}\epsilon_{jlm}\omega_l \langle s_k s_m\rangle \right]
\end{eqnarray}
This average magnetization dynamics relies on higher order correlation functions of noise and spin that have to be derived. 
By  applying the Shapiro-Loginov method \cite{shapiro1978formulae,berdichevsky1999stochastic} to the nine components of the $\langle \tilde{\omega}_i s_j \rangle$ matrix at the same time, we obtain the equation:
\begin{equation}
\frac{ \partial\langle \tilde{\omega}_i s_j \rangle}{\partial t}= \left\langle \tilde{\omega}_i \frac{\partial{s}_j}{\partial t} \right\rangle -\frac{1}{\tau}\left\langle \tilde{\omega}_i s_j \right\rangle
\label{Shapiro}
\end{equation}
Injecting the right-hand side of Eq.(\ref{LLG}) in Eq.(\ref{Shapiro}), equations for  the second-order moments $\langle \tilde{\omega}_i s_j\rangle$ are obtained.

The equations for the moments $\langle s_i s_j\rangle$ are now required. Assuming that the following identity
\begin{equation}
\frac{\partial \langle s_i s_j\rangle}{\partial t} = \left\langle \frac{\partial {s}_i}{\partial t} s_j\right\rangle+\left\langle s_i \frac{\partial{s}_j}{\partial t}\right\rangle
\label{partials}
\end{equation}
holds, the time derivatives of ${\partial s_i}/{\partial t}$ are replaced by the right-hand side of Eq. (\ref{LLG}) and so forth.

This leads to an additional set of nine equations for the second-order moments of the microscopic spin degrees of freedom,  that are closely related to those written by Garanin \emph{et al.} (see Eq.(6) in ref. \cite{garanin1990dynamics}) and Gracia-Palacios \emph{et al.} (see Eq.(2.10) in ref. \cite{garcia1998langevin}).

The complete system of equations, therefore, takes the following form:
\begin{eqnarray}
\frac{ \partial \langle s_i\rangle}{\partial t}   
&=&\frac{1}{1+\lambda^2} \left[  \epsilon_{ijk}\omega_j \langle s_k\rangle+ \epsilon_{ijk} \langle\tilde{\omega}_j s_k\rangle\right. \nonumber\\  
                                  ~&~&\left.+\lambda \epsilon_{ijk}\epsilon_{jlm}\omega_l \langle s_k s_m\rangle\right]\\ 
\frac{\partial\langle\tilde{\omega}_i s_j\rangle}{\partial t} 
&=&-\frac{1}{\tau} \langle \tilde{\omega}_i s_j \rangle+\frac{1}{1+\lambda^2} \left[\epsilon_{jkl}\omega_k \langle\tilde{\omega}_i s_l\rangle  \right. \nonumber \\ 
                                  ~&~&+\epsilon_{jkl}\langle\tilde{\omega}_i  \tilde{\omega}_k s_l\rangle  \nonumber\\
                                  ~&~&+\left.\lambda \epsilon_{jkl} \epsilon_{lmn}\omega_m \langle\tilde{\omega}_i s_k s_n\rangle \right]\\
\frac{\partial\langle s_i s_j \rangle }{\partial t}      
&=&\frac{1}{1+\lambda^2}\epsilon_{jkl}\left(\omega_k \langle s_i s_l\rangle + \langle \tilde{\omega}_k s_i  s_l\rangle\right.\nonumber\\
                                  ~&~&\left.-\lambda\epsilon_{lmn}\omega_m \langle s_i s_l s_n \rangle \right)+ \left(i \leftrightarrow j\right)
\label{BBGKY}
\end{eqnarray}

Due to the non-linearity of the sLLG equation, an infinite hierarchy arises \cite{nicolis1998closing,adomian1971closure}: these equations for  the second-order moments depend 
on third-order moments, and so on. In order to solve the system, closure relations need to be found, that express the third-order moments in terms of the second order and first order moments. 
These relations can be  deduced from the definition of the corresponding cumulant moments. 
 
The double bracket notation  $\langle \langle.\rangle\rangle$ stands for the cumulant of the stochastic variables \cite{van1992stochastic}, and for any stochastic vector $\vec{x}$, one has:
\begin{eqnarray}
\langle\langle x_i x_j x_k\rangle\rangle &=& \langle x_i x_j x_k\rangle - \langle x_i \rangle\langle x_j x_k\rangle \nonumber \\
                                        ~&~&- \langle x_j\rangle \langle x_i x_k\rangle - \langle x_k \rangle \langle x_i x_j\rangle \nonumber\\
                                        ~&~&  + 2 \langle x_i \rangle \langle x_j \rangle \langle x_k \rangle 
\label{Cumulants}
\end{eqnarray}
Assuming that the third-order cumulants vanish for each stochastic variable (meaning that the corresponding distribution is assumed  Gaussian), and applying the 
relation (\ref{Cumulants}), the following closure relations are deduced:
\begin{eqnarray}
\langle\tilde{\omega}_i s_j s_k\rangle             &=&  \langle s_j\rangle \langle\tilde{\omega}_i s_k \rangle + \langle s_k\rangle \langle\tilde{\omega}_i s_j \rangle \label{closure1}\\
\langle\tilde{\omega}_i \tilde{\omega}_k s_l\rangle&=&\frac{D}{\tau}\delta_{ik} \langle s_l\rangle \label{closure2}\\
\langle s_i s_j s_k\rangle                         &=& \langle s_i\rangle\langle s_j s_k\rangle + \langle s_j\rangle\langle s_i s_k\rangle \nonumber\\
                                                   &~& + \langle s_k\rangle \langle s_i s_j\rangle -2\langle s_i\rangle\langle s_j\rangle\langle s_k\rangle \label{closure3}\
\label{KY}
\end{eqnarray}
The second closure relation arises from the properties of the noise for the same time: indeed, one has $\langle \tilde{\omega}_i \rangle=0$ and 
$\langle\tilde{\omega}_i\tilde{\omega}_j \rangle = \frac{D}{\tau}\delta_{ij}$ when $\tilde{\omega}_i$ and $\tilde{\omega}_j$ are considered at the same time.
Introducing those relations in the system (\ref{BBGKY}), we finally have:
\begin{eqnarray}
\frac{\partial \langle s_i\rangle}{\partial t}   &=&\frac{1}{1+\lambda^2} \left[ \epsilon_{ijk}\omega_j \langle s_k\rangle+ \epsilon_{ijk}\langle \tilde{\omega}_j s_k\rangle \right.\nonumber\\
                                 ~&~&-\left.\lambda \epsilon_{ijk}\epsilon_{jlm}\omega_l  \langle s_k s_m \rangle \right]\\
\frac{\partial \langle \tilde{\omega}_i s_j\rangle}{\partial t}&=&-\frac{1}{\tau} \langle \tilde{\omega}_i s_j\rangle+\frac{1}{1+\lambda^2} \left[\epsilon_{jkl}\omega_k  \langle \tilde{\omega}_i s_l\rangle  \right.  \nonumber\\
                                 ~&~&+\frac{D}{\tau}\epsilon_{jil}\langle s_l\rangle\nonumber\\
                                 ~&~&-\lambda\epsilon_{jkl}\epsilon_{kmn}\omega_m \left( \langle s_l\rangle \langle \tilde{\omega}_i s_n\rangle   \right.\nonumber\\
                                 ~&~&+\left.\left. \langle s_n\rangle  \langle \tilde{\omega}_i s_l\rangle  \right)\right]\\
\frac{\partial \langle s_i s_j\rangle}{\partial t}&=&\frac{1}{1+\lambda^2} \left[ \epsilon_{jkl}\omega_k \langle s_i s_l\rangle  \right.\nonumber\\
                                 ~&~&+ \epsilon_{jkl} \left( \langle s_i\rangle \langle \tilde{\omega}_k s_l\rangle +\langle s_l\rangle  \langle \tilde{\omega}_k s_i\rangle  \right) \nonumber\\
                                 ~&~&- \lambda\epsilon_{jkl}\epsilon_{kmn}\omega_m \left( \langle s_i\rangle  \langle s_l s_n\rangle \right.\nonumber\\
                                 ~&~&+ \langle s_l\rangle \langle s_i s_n\rangle  +\langle s_n\rangle \langle s_i s_l\rangle  \nonumber\\
                                 ~&~&-\left.\left.2 \langle s_i\rangle\langle s_l\rangle\langle s_n\rangle\right)\right] + \left(i \leftrightarrow j\right)
\label{AveragedSystem}
\end{eqnarray}
This set of twenty-one differential equations is now closed and can be integrated simultaneously to describe the averaged dynamics of the effective magnetic system connected to a bath.
It should be noted  that the last two equations  involve both, longitudinal and transverse, forms of damping.
In consequence thereof, the first equation, that describes  the evolution of the averaged magnetization, has a Landau-Lifshitz-Bloch form, with both, transverse and longitudinal, damping terms. 

In the following sections, this model will be called the dynamical-Landau-Lifshitz-Bloch (dLLB) model.

\section{Numerical experiments}
Numerical experiments are performed in order to test the consistency of our closure hypothesis. 

We first focus on the limiting case of small correlation time, i.e. the vicinity of the white-noise limit.
Indeed, if $\tau$ is very short (the same order of magnitude as our integration time), the associated relaxation can no longer be resolved, and 
loses its physical meaning. Under those conditions, the dynamics is Markovian, and our dLLB model is expected to match the results obtained from an averaged Markovian stochastic magnetization dynamics. 

In order to realize these comparisons, a reference model is built from  the sLLG (Eq.~(\ref{LLG})) with a white-noise random vector.  
One thousand realizations of this stochastic equation with the same initial conditions are generated, and a statistical average is taken over them. The dLLG system is solved by using an eight-order Runge-Kutta integration scheme with adaptive steps constructed by the Jacobian of the whole system.  

\begin{figure}[thp]
\resizebox{0.95\columnwidth}{!}{\includegraphics{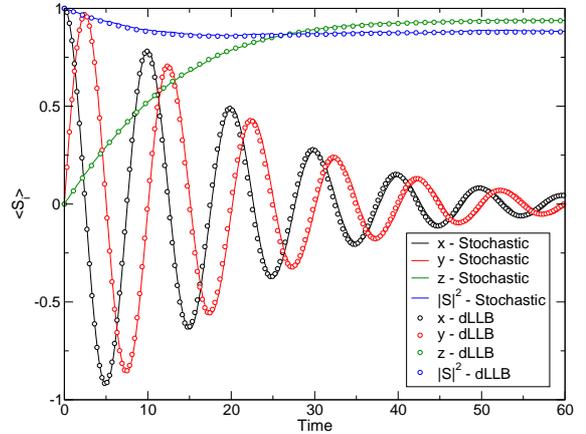}}
\caption{Relaxation of the averaged magnetization dynamics about an external field in the $z$-direction and in the case of small correlation time for the noise. 
The solid lines plot the stochastic case (1000 averaged repetitions), dots are for the dLLB model. Simulations parameters: \{$D=1.6.10^{-2}$; $\lambda=0.1$; $\tau=10^{-3}$; 
$\vec{\omega}=(0,0,\pi/5)$ \}. Initial conditions:$\vec{s}(0)=\left(1,0,0 \right)$.   \label{Fig1}}
\end{figure}

Fig.~(\ref{Fig1}) shows the  average, obtained from   the sLLG equation, and the dLLB model for the same value of the noise amplitude $D.$
It appears that the magnetization dynamics obtained with our dLLB model fits very well the averaged sLLG results in the case of small correlation time for the noise.
Moreover, an essential feature of the Fig.~(\ref{Fig1}) is that our model does behave like an LLB equation and can describe the variation of the magnetization norm, that does not remain constant. This is due to the longitudinal damping component that is generated by  the average effects in presence of the  noise.

\begin{figure}[thp]
\resizebox{0.95\columnwidth}{!}{\includegraphics{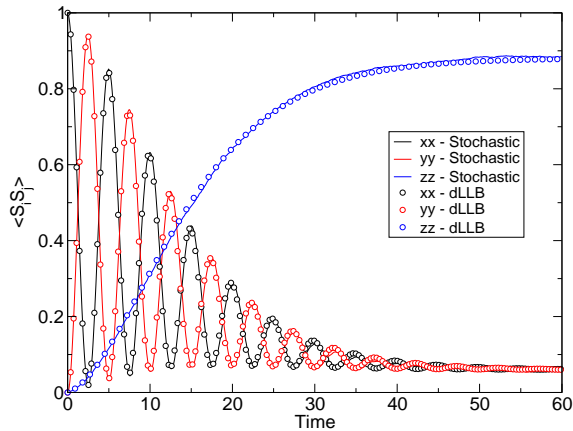}}
\caption{Relaxation of the diagonal components of the correlation tensor $\langle s_is_j\rangle$. Again,the solid lines plot the stochastic case (1000 averaged repetitions) and the dots are for 
the dLLB model. Initial conditions: $\langle s_1(0)^2\rangle=1$, and $\langle s_i(0) s_j(0)\rangle=0~ \forall (i,j) \neq (1,1) $ . \label{Fig2}}
\end{figure}

The second order moments,  also,  play a major role in the evolution of the averaged magnetization dynamic equations, and the diagonal components of the tensor 
$\langle s_is_j\rangle$ are  plotted in Fig.~(\ref{Fig2}).\\
Good agreement is also found between the two  models for the diagonal components of the correlation tensor $\langle s_is_j\rangle$.
Those results allows us to test the validity of the equality Eq.~(\ref{partials}), and the closure relations Eqs.~(\ref{closure1},\ref{closure2},\ref{closure3}).

\section{Conclusion}
Statistical average and the application of the Shapiro-Loginov method  allow us to derive  a new formalism for the magnetization dynamics 
when a colored form for the noise is appropriate. 
Our model has proven to be effective for the description of an average magnetization dynamics when small values of the time-correlation $\tau$ are considered.
In this particular case, the results of Markovian dynamics are recovered 
and very good agreement was found between our dLLB set of equation and an averaged set of LLG equation with a white-noise process.
Moreover, this work demonstrates that the application of statistical average to a magnetic system 
in contact with a thermal bath leads to the direct appearance  of a longitudinal form of damping, and to equations consistent with the Landau-Lifshitz-Gilbert model.
In future work, the effect of longer  correlation times for the noise will be studied in detail, and comparison between our dLLB models and atomistic 
magnetization dynamics in more complex situations (exchange interaction, anisotropy, superparamagnetism, etc.) will be presented. \\

JT acknowledges financial support through  a joint doctoral fellowship ``R\'egion Centre-CEA''.  \\

\bibliographystyle{unsrt}
\bibliography{LLB}
\end{document}